# Hydrogen Spectra, Molecular Association and Orbital Radii in the Solar System

James C. Lombardi

Physics Department, Allegheny College, Meadville PA, 16335

E-mail:  [jlombard@allegheny.edu](mailto:jlombard@allegheny.edu)

**ABSTRACT**

A relationship between the average orbital radii of the planets and their satellites in the solar system and the spectra of atomic and molecular hydrogen is identified and investigated.  In this model, stimulated radiative association resonances develop early on in the disk of the protosun that cause the disk to cool at only certain radii, with each radius depending on a specific photon energy in the atomic hydrogen spectrum.  The planets then evolve from the relatively cool rings that are formed.  Similar activity occurs in the formation of the satellite systems of the giant planets.  The present investigation deals with the mechanism that generates rings from which the planets are formed.  It does not deal with the evolution of the rings into planets.  Many characteristics of the solar system are explained including the sizes of the orbital radii of the planets and their satellites, the tilt of Uranus's axis, the positions of the asteroid and Kuiper belts, the source of the scattered Kuiper belt objects, the positions of Saturn's main rings and the rings of Uranus, Jupiter, and Neptune.  It also shows that a commonality exists in the structures of the solar system and the planetary systems that can be attributed to the common process that initiated their evolution.

1. Introduction

The problem dealing with the sizes of planetary orbits has intrigued many investigators (e.g., Nieto 1972; Lissauer and Cuzzi 1985). Some calculations involving gravitational resonances (e.g., Torbett et al. 1982) have done well to match the over 200 year-old Titius-Bode law for the orbital radii of the planets. The present investigation sets forth a model that explains not only the orbital radii in the solar system but also in the systems of the giant gas planets. This model also deals with resonance; however the resonance involves the interaction of light with atoms and molecules in the primordial disks of the protosun and protoplanets. It is based on a connection that exists between the energies of photons in the spectra of atomic and molecular hydrogen and the average orbital radii of the planets and their satellites in the solar system. A reasonable explanation for the connection is given by an extension of the discussion by Stancil and Dalgarno (1997a, 1997b) and Zygelman et al. (1998) concerning stimulated radiative association of atoms and ions to form molecules. Stancil, Dalgarno and Zygelman, (SDZ) considered processes of the form

$$A + B \rightarrow AB + \nu, \quad \text{(without stimulation)} \quad (1)$$

and
$$A + B + \nu \rightarrow AB + 2\nu, \quad \text{(with stimulation)} \quad (2)$$

where AB is the molecule formed by the association of the atoms (or atom and ion) A and B. In Eqs. (1) and (2) each $\nu$ represents a photon with energy $E_P$. Furthermore,

$$E_P = E_C + E_B, \quad (3)$$

where the temperature dependent collision energy $E_C$ is the thermal energy lost in the association and the binding energy $E_B$ corresponds to the state in which the molecule is formed.

SDZ considered stimulated radiative associations occurring in the presence of a blackbody radiation field and resulting in molecules such as $HeH^+$, LiH, and the deuterated hydrogen molecule HD. In the present model, stimulated radiative association occurs in the presence of photons with specific energies, rather than by a smoothly varying blackbody distribution. The model deals with a protosun (or gaseous protoplanet) consisting of a central core and a circulating disk with the temperature of the disk varying with the radial coordinate $r$. In the case of the solar system, discussed in § 2, radiation emitted from highly excited hydrogen atoms in the disk or the core is directed or redirected by scattering into the plane of the disk. This radiation interacts with colliding atoms, initiating processes described by Eq. (2). Furthermore, Eq. (2) involves photons stimulating more photons, thus supplying the possibility of resonance similar to that of a laser. The temperature is constant along rings in the disk's plane, and resonances are initiated in rings where the collision energy is such that Eq. (3) is satisfied. As indicated by Eq. (2), the stimulated radiative association process causes kinetic energy to be lost to photon energy and the disk is therefore cooled in just certain rings. In this model, once conditions are met for a resonance to occur, it will persist at the same radius even though the temperature and average kinetic energy of the atoms at that radius continually drop. At any temperature there will always be pairs of colliding atoms with kinetic energy values that cause Eq. (3) to be satisfied.

Presumably the planets are ultimately formed from material that collects in the cool rings. This investigation does not deal with the creation of the planets from rings but only with their placement relative to the Sun's center. Each ring corresponds to a certain photon energy; hence the connection between the hydrogen spectrum and average orbital radii of the planets.

The resonance proposed here is characterized by the (E)nhancement of (IN)elastic collisions by (S)timulated (E)mission of (R)adiation, referred to as EINSER in this paper. EINSER involves the process described by Eq. (2) but further involves the resonance described above, the enhancement of inelastic collisions and the concurrent cooling of the gas. Findings in § 7 indicate that for the Sun and Saturn the molecules formed as a product of the resonance are $H_2^+$ molecular ions. Therefore for the Sun and Saturn Eq. (2) is the specific relationship

$$H + H^+ + \nu \rightarrow H_2^+ + 2\nu. \qquad (4)$$

Eq. (4) indicates that as long as the $H_2^+$ is always created in the same state, the relative structure in the solar system depends on the energies of the photons present in the disk, and not on the energy levels in $H_2^+$. We can think of the second photon appearing on the right side of Eq. (4) as being created from the relative kinetic energy (thermal energy) of the H and $H^+$ that is lost in the molecular association process.

This model predicts the formation of the asteroid belt (§ 2.2) and the classical Kuiper belt (§ 2.3). A third belt (§ 2.4) is predicted to have existed in the region of Uranus and Neptune. No belt is presently observed in this zone, but such a belt could have been the source of the observed scattered Kuiper belt objects. A fourth belt (§ 2.4) is predicted well within the orbit of Mercury. In § 3-6 relationships are found to exist between the spectra of atomic and molecular hydrogen and the average orbital radii of the satellites of planets. These relationships indicate a commonality in ring production existed during the creation of the solar and planetary systems.

Support for the present hypotheses is found in the connections that are identified throughout this paper as existing between orbital radii in the solar system and energy levels in atoms and molecules. For instance, (1) the analysis presented in §7 shows a link between structure in Saturn's ring system and binding energies of states in the $H_2^+$ molecular ion. To focus on this aspect, § 2, § 2.1 and § 2.2 should be read for background material and then § 3, § 3.1 and § 7. (2) A similarity exists between the positions of Uranus's rings and the pattern of peaks seen in part of the $H_2$ spectrum (§ 4). (3) The positions of the asteroid and Kuiper belts in the solar system are related to structure in the hydrogen atom (§ 2.2 and § 2.3). (4) The assumed relationships between orbital radii and energy levels in atoms and molecules leads to the determination that scaled temperature distributions of the protodisks of the giant planets are essentially the same (Fig. 6). Evidence supporting the present model builds throughout this paper and the combination of all the results compliment each other and help validate the underlying hypotheses.

## 2. The Solar System

We begin by searching for the photon energies $E_P$ that are associated with the orbital radii of the planets. The collision energy $E_C$ depends on the disk temperature and, assuming temperature is smoothly varying with distance from the protosun's center, the correct set of $E_P$'s could produce a smoothly varying graph of $E_P$ versus average orbital radius. To find this set, we consider two adjacent EINSER resonances that ultimately result in the creation of two neighboring planets. Assuming the binding energy $E_B$ is the same for each resonance, Eq. (3) gives $\Delta E_P = \Delta E_C$, where $\Delta E_P$ is the difference in photon energies and $\Delta E_C$ is the difference between the collision energies associated with each resonance. Basic mechanics can be used to show that on average the collision energy in a molecular association process is equal to the average kinetic energy of the particles that are merging. Therefore, upon initiation of resonance $\Delta E_P = 3/2\, k\Delta T$, where $\Delta T$ is the difference in temperature between the two vicinities where resonances are occurring and k is Boltzmann's constant. For example, the temperature difference $\Delta T$ between the positions of Jupiter and Saturn in the solar disk is about 50-70 K (Cameron 1978), yielding a $\Delta E_P$ value in the range 52-73 cm$^{-1}$. Assuming the photons in the disk are predominantly those found in the spectrum of atomic hydrogen and also assuming resonances for neighboring planets correspond to successive $E_P$'s in one of the hydrogen photon energy series (Lyman, Balmer etc.), the search is limited to regions of each series where the difference in photon energies is about 52-73 cm$^{-1}$.

The following formula holds for photon energies in the hydrogen spectrum.

$$E_P(n_f, n_i) = 109737 \text{ cm}^{-1} (1/n_f^2 - 1/n_i^2), \quad (n_f = 1,2,3\cdots, n_i = (n_f+1), (n_f+2)\cdots), \quad (5)$$

where the quantum numbers $n_f$ and $n_i$ are for the final and initial states associated with the emission of a photon from a hydrogen atom. Using Eq. (5), a region of the hydrogen spectrum from $E_P(8,12) = 952.58$ cm$^{-1}$ to $E_P(8,22) = 1487.9$ cm$^{-1}$ is found to result both in a smoothly varying graph of $E_P$ versus observed average orbital radii and in a $\Delta E_P$ for Jupiter and Saturn of 59 cm$^{-1}$. This set of photon energies not only includes $E_P$'s for which $n_f = 8$ but a few other values of $E_P$ for which $n_f = 5, 6$, and 7, and many for which $n_f = 9$ and 10. All values of $E_P$ that fall in the region of interest (from Mercury to Pluto) have been explicitly listed in Table 1 except for the two series with $n_f = 9$ and 10. For these, only the series limit is listed. These limits and the limit with $n_f = 11$ are apparently associated with inner radii of belts in the solar system as will be discussed in § 2.2 - 2.4. Other schemes matching $E_P$'s with orbital radii in the solar system have been investigated and will be briefly discussed in § 10. The only suitable one that could be found is the scheme given in Table 1.

## 2.1 The Planets and Largest Asteroids

In this discussion $r$ is the radial coordinate measured from the center of the Sun and $R$ is the average orbital radius (length of the semi-major axis of the orbit) of a planet. The value of $R$ for each of the planets is listed in Table 1 and matched with its corresponding $E_P$. Fig. 1 is a plot of $E_P$ versus $R$ for the solar system. When this table and figure were first generated there were four fairly close $E_P$'s (1341.2, 1334.9, 1333.6, and 1332.6 cm$^{-1}$) that were not matched with average orbital radii of planets. Ceres, Pallas, and Vesta are the three most massive asteroids (Hilton J. L. (2003) In *U. S. Naval Observatory Ephemerides of the Largest Asteroids*, http://aa.usno.navy.mil/ephemerides/ asteroid//astr_alm/asteroid_ephemeredes.html) and it is therefore reasonable to attempt associating them with available $E_P$'s. The attempt is successful as can be seen in Fig. 1 since the resulting three points fall on the curve that connects the other data points. To determine the asteroid(s) with which the fourth $E_P$ should be associated, a linear fit is made for the Ceres, Pallas, and Vesta data. This fit and the fourth value in question, $E_P(8,17) = 1334.9$ cm$^{-1}$ is used to predict the fourth asteroid's orbital radius to be 2.68 AU. The IRAS data for the asteroid diameters (Bowell E. of Lowell Observatory (2003) In *The Asteroid Orbital Elements Database* http://www.lowell.edu/users/elgb/) indicates that among the 16 largest asteroids there are just 6 with orbital radii less than 2.9 AU; the three largest asteroids (Ceres, Pallas, and Vesta) that are already accounted for, Eunomia at 2.645 AU, Juno at 2.668 AU, and Bamberga at 2.683 AU. The last three have orbital radii all near the predicted value of 2.68 AU and their average orbital radius is 2.67 AU. Therefore the last $E_P$ is finally matched with this average value under the assumption that the three asteroids evolved from the same resonance and resulting ring. In this model the rest of the asteroid belt evolves from EINSER resonances corresponding to photons in the $n_f = 9$ series in Eq. (5) as will be discussed in § 2.2.

The following empirical formula is found by adjusting four parameters to fit the data for planets from Venus to Pluto in Table 1.

$$E_P = (C_0/r^p + C'r^2 + B_0), \qquad (6)$$

where $p = 0.108$, $C_0 = 774$, $C' = -0.13$, $B_0 = 637$ cm$^{-1}$, $r$ is measured in AU's from the Sun's center and energy is in wave numbers. The curve plotted in Fig. 1 is a graph of Eq. (6). Similar empirical formulas will be seen to hold for each of the satellite systems of the planets Saturn, Uranus, Jupiter, and Neptune. These formulas serve as convenient approximations for the relationships between $R$'s and $E_P$'s.

One has the tendency to equate $E_B$ from Eq. (3) with the constant 637 cm$^{-1}$ in Eq. (6) and $E_C$ from Eq. (3) with the radially dependent terms in Eq. (6). But the collision energy $E_C$, which certainly is a function of $r$ since it depends on the radially dependent temperature, might also have a component in it that together with the actual $E_B$ would produce the constant term in Eq. (6). On the other hand it is helpful to think of the graph in Fig. 1 as having a shape, albeit shifted vertically by some constant, which is at least roughly proportional to the temperature distribution in the protosun's disk. For the sake of discussion the first term on the right side of Eq. (6) will be referred to as the effective collision energy function and a particular value of the function as an effective collision energy $E_C'$. The $C'r^2$ term in that equation is necessary for the analysis of the solar system data but not for the analysis of the data associated with the planets. This term possibly helps to model the effects on the positions of the planets due to the change of mass over time in the Sun's disk or in the Sun itself. Also, the value of the parameter $p$ is likely to be affected by the evolution of the mass in the Sun and its disk.

## 2.2 The Asteroid Belt

The series $E_P(n_f=9, n_i=10,\cdots,\infty)$ of photon energies in the hydrogen spectrum overlap the $E_P$'s given in Table 1. The separation between photon energies gets smaller for larger values of $n_i$. In the present model, photons with close energies create a continuous region of EINSER resonance within the boundaries of the asteroid belt (or within the boundaries of any of Saturn's main rings). Since $r$ is decreasing as $E_P$ increases in Eq. (6), the limiting value of very closely spaced $E_P$'s ($E_P(9,\infty) = 1354.8$ cm$^{-1}$) corresponds to the inner edge of very closely spaced resonances. It is assumed here that many asteroids eventually evolve to form a belt in the relatively cool zone that results. Equation (6) and the value of $E_P(9,\infty)$ are used to determine the radius of the inner edge of the belt to be 2.0 AU. This value (indicated by the triangle on the curve in Fig. 1) is in excellent agreement with the known inner radius of the asteroid belt near 2.0 AU (Clark 1999).

The asteroid belt extends out to an outer edge radius of about 4 AU (Clark 1999) with gaps associated with the gravitational interaction of asteroids with Jupiter. This radius and Eq. (6) can be used to determine a value of $E_P$ near $E_P(9,45) = 1300.6$ cm$^{-1}$ corresponding to the belt's outer edge. So, in this model the resonances that ultimately produced most of the asteroids in the asteroid belt correspond to the range $E_P(9,45)$ to $E_P(9,\infty)$. As will be shown in § 3.1, this is the same range of $E_P$'s that fits the E ring of Saturn.

Consideration should be given to why there is an outer edge cut off to the asteroid belt. The emission intensities in atomic hydrogen are generally lower for larger values of $n_f$ in Eq. (5) (National Institute of Standard and Technology Spectra Data Base (2003). http://physlab2.nist.gov/cgi-bin/AtData/main_asd). Possibly the $n_f = 9$ transitions in hydrogen have intensities that are not strong enough to cause resonance unless they correspond to EINSER resonances that spatially overlap and therefore mutually support each other. Presumably transitions with $n_i \sim 46$ and lower do not cause overlap.

The prediction of the asteroid belt and the accurate calculation of the position of its inner edge are important successes of the present model. Similar successes are discussed in § 2.3 for the Kuiper belt and § 3.1 and § 7 for Saturn's rings.

## 2.3 The Kuiper Belt

Three other belts in the solar system are predicted in this model. One of them has an inner edge corresponding to the series limit $E_P(11,\infty) = 906.9$ cm$^{-1}$. The inner edge radius of 43.6 AU, indicated by the square in Fig. 1, is calculated using this limit and Eq. (6). In the case of the asteroid belt it was previously determined that the outer edge radius corresponded to the value of $n_i = 45$. Assuming the same value of $n_i$ for the outer edge of the belt, an estimate of the outer edge position can be determined. Using $E_P(11,45) = 852.7$ cm$^{-1}$ in Eq. (6) the outer edge radius is calculated to be 47.6 AU. Therefore a belt is predicted to exist in the range from 43.6 to approximately 47.6 AU. The orbital elements of many Kuiper Belt Objects (KBO's) have been measured (e.g., Jewitt 1999). The classical KBO's (their orbits have small eccentricities) exist in a belt characterized by objects with semi-major axes mainly between 42 and 47 AU. The position of this predicted second belt corresponds very well with the position of the main component of the observed Kuiper belt.

## 2.4 Other Belts Around the Sun

The present model predicts two more belts around the Sun. One is in the Uranus-Neptune zone and another is near the Sun. The inner edge radius of the belt in the vicinity of Uranus and Neptune is determined as before by substituting the series limit $E_P(10,\infty) = 1097.4$ cm$^{-1}$ into Eq. (6). The resulting inner edge radius 25.6 AU, is indicated by the diamond in Fig. 1. If $n_i = 45$ for the outer edge of this belt as well, then $E_P(10,45) = 1043.2$ cm$^{-1}$ and the outer edge radius is calculated to be 31.3 AU. The value of the outer radius is not firmly established because the $n_i$ value used to determine it is simply chosen to be the same as the one determined from the asteroid belt. But it does help to establish an estimate of the belt's radial extent (25.6 to 31.3 AU). The inner radius of the belt lies between the orbital radii of Uranus and Neptune, and the outer radius just beyond Neptune's average orbital radius at 30.0 AU. No belt has been observed to exist presently in the Uranus-Neptune zone. However, Jewitt et al. (1998) suggest the observed scattered KBO's (their orbits have large eccentricities and inclinations) were scattered out of the Uranus-Neptune zone by Neptune when the solar system was young. The third belt could therefore be the source of the scattered KBO's.

Most of the planets correspond to $E_P$'s with $n_f = 8$. This series of $E_P$'s predicts a belt closer to the Sun than Mercury. The inner edge radius of this ring corresponds to $E_P(8,\infty) = 1714.7$ cm$^{-1}$. Using this value for the inner edge of a belt and $E_P(8,45) = 1660.5$ cm$^{-1}$ for the outer edge, the prediction for the radial extent of a narrow fourth belt is 0.047 to 0.075 AU. The accuracy of the predicted belt position depends on the accuracy of Eq. (6) when extrapolated to radii less than the orbital radius of Mercury. Even though no belt has been observed close to the Sun, the series limit $E_P(8,\infty) = 1714.7$ cm$^{-1}$ does have significance in Saturn's system where it will be seen to correspond to the inner edge of Saturn's D ring. The results for the four solar system belts are listed in Table 2.

## 2.5 Mercury, Uranus, and Pluto

In Table 1 and Fig. 1, three $E_P$'s have been tentatively grouped for Mercury. Possibly, closely spaced EINSER generated rings can evolve to make protoplanets that collide and combine to make larger planets. This could have occurred in the case of Mercury, or maybe one or two of the three predicted protoplanets near Mercury's orbit were scattered into the Sun or became one or two of the many asteroids that orbit in the inner part of the solar system. A collision between one of them and Venus could explain Venus's retrograde rotation.

Two rings and therefore two protoplanets are predicted in the vicinity of Uranus and are assumed to have evolved into the single planet. If the protoplanets collided to form Uranus, this could explain its retrograde rotation and nearly 90$^0$ tilt of its axis of rotation (Korycansky et al. 1990; Slattery et al. 1992). Just before the collision the angular momentum vector associated with the objects' mutual approach could have been roughly parallel to the orbital plane. The resulting composite planet would have its axis of rotation tilted severely.

The semi-major axis of Pluto's orbit (39.5 AU) was used with other data in the fit to find the parameters in Eq. (6). Actually there are many KBO's with nearly the same average orbital radius as Pluto's. They are orbiting just inside the main belt of KBO's and have recently been labeled Plutinos (Jewitt 1999). In the present model the proximity of the Plutinos to each other is attributed to their formation occurring in the same relatively cool region created by the EINSER resonance associated with the photon energy $E_P(8,12) = 952.6$ cm$^{-1}$.

### 3. Saturn's Satellite System

The EINSER model can also be used to analyze Saturn's system. In this analysis and in the analyses of the systems for the other giant planets only satellites that move with nearly circular orbits and with low inclinations of their orbital planes are considered. First a series of photon energies $E_P$'s that matches with Saturn's satellite orbital radii must be found. Note that near the inner edge of the Sun's asteroid belt is the cluster of six large asteroids discussed in § 2.1. Five of these have nearly the same orbital radius. The orbital radius of the sixth asteroid, Vesta, is somewhat smaller. Similarly, near the inner edge of Saturn's E-ring is a cluster of four satellites (Enceladus, Tethys, Telesto, and Calypso). The last three have nearly the same orbital radius, and Enceladus's orbital radius is somewhat smaller. This pattern leads to matching Enceladus's orbital radius with the same photon energy ($E_P(5,6) = 1341.2$ cm$^{-1}$) that corresponds to Vesta's orbital radius. With this as the key, the scheme indicated in Table 1 relating orbital radii $R$ and $E_P$'s is established for Saturn including the data for the inner edge of Saturn's D ring, the closest ring to Saturn. It also lists scaled radii for Saturn's satellites that will be discussed in § 6.

Fig. 2 is a plot of the data found in Table 1 for Saturn. The solid curve in that figure is a graph of the

fitted equation

$$E_P = (8032/(r - r_0)^p + B_0), \qquad (7)$$

where $p = 0.196$, $r_0 = 37,800$ km, $B_0 = 637$ cm$^{-1}$, and $r$ is in km from Saturn's center. The points in the dip centered around $r = 2 \times 10^5$ km were not included in the fit. Preliminary fits to determine parameters for the Sun and Saturn in Eqs. (6) and (7) produce $B_0$'s that are approximately the same in each equation. Therefore $B_0$ is constrained to be the same value in the two fits and found to be 637 cm$^{-1}$. The term for the effective collision energy function in Eq. (7) diverges at the radius $r_0 = 37,800$ km. This divergence should not be a concern, as Eq. (7) is meant to describe the effective collision energy function only for $r$ considerably larger than $r_0$.

### 3.1 Saturn's D and E rings.

The inner edge radius of Saturn's D ring, 66,000 km (Showalter 1996), was used in the fit to determine Eq. (7) with the corresponding value of $E_P(8,\infty) = 1714.7$ cm$^{-1}$. In the cases of the asteroid and the Kuiper belts, the outer edges were each found to correspond to $n_i = 45$. Assuming the same value of $n_i$ for the D ring, its outer edge radius is estimated to be $7.45 \times 10^5$ km by substituting $E_P(8,45) = 1660.5$ cm$^{-1}$ into Eq. (7). This matches the observed value of $7.451 \times 10^5$ km (Cuzzi et al. 1984).

As discussed in § 2.2 the asteroid belt in the solar system is successfully predicted by the present model, with its inner and outer edge radii corresponding to $E_P(9,\infty) = 1354.8$ cm$^{-1}$ and $E_P(9,45) = 1300.6$ cm$^{-1}$ respectively. The E ring of Saturn is also associated with the identical range of $E_P$'s. Using $E_P(9,\infty) = 1354.8$ cm$^{-1}$ to interpolate between points of nearby satellites the inner edge radius $2.13 \times 10^5$ km is calculated. The diamond in Fig. 2 indicates the calculated inner edge. Similarly, the outer edge radius calculation is $3.60 \times 10^5$ km (the square in Fig. 2). These calculated radii are in good agreement with the observed inner and outer radii of Saturn's E ring near $1.8 \times 10^5$ and $3.6 \times 10^5$ km respectively (de Pater et al. 1996; Bauer et al. 1997).

## 4. Uranus's Rings and Satellites

The orbital radii in the satellite systems of Uranus, Jupiter, and Neptune cannot be reasonably matched with a set of $E_P$'s in the spectrum of the hydrogen atom. However they can be matched with a set of $E_P$'s that exist in the $H_2$ infra-red emission spectrum as seen in reflection nebulae. It would be impossible to identify this set as the correct one if not for Uranus's rings, as will be discussed in this section.

Martini et al. (1999) have measured the spectra of four reflection nebulae, NGC 1333, NGC 2023, NGC 2068, and NGC 7023. Fig. 3 is a graph of a theoretical spectrum (provided by Paul Martini) that Martini et al. generated by smoothing a model spectrum determined by Draine and Bertoldi (1996) to the resolution of their detector. This theoretical spectrum matches the experimental results of Martini et al. very well. The pattern seen in the positions of Uranus's rings (Elliot 1979) and satellites Ophelia and Cordelia is very similar to the pattern of the 2-micron region of peaks from 1.85 to 2.29 μm (5400 to 4370 cm$^{-1}$) seen in the $H_2$ spectrum. In Fig. 3 each peak in the pattern is labeled with the name of the Uranian ring or satellite with which it is associated. This leads to exploration of the possibility that radiation seen in reflection nebulae is also the radiation that produced EINSER resonance in Uranus's disk.

From the theoretical work of Black and van Dishoeck (1987), it is seen that the spectral peaks associated with Uranus's rings correspond to S-branch transitions in $H_2$ that have intensities greater than about 6 % of the intensity of the strongest line. In Fig. 3 the three small peaks between the ones associated with Ophelia and the ε ring belong to Q and O transitions. All the labeled peaks in Fig. 3 correspond to S transitions. Half of Table 3 lists Uranus's rings and satellites with orbital radii less than 54,000 km with their corresponding $E_P$'s taken from Black and van Dishoeck. Note, in contrast to Fig. 1 and 2 for the Sun and Saturn, the values of $E_P$ increase rather than decrease with $r$ out to about 54,000 km. Since each $E_P$ and its corresponding collision energy differ by a constant, the collision energy function and therefore the temperature also increase with increasing $r$ throughout the region of the rings. But the temperature cannot continue to increase indefinitely and there eventually will be a maximum in the temperature distribution followed by a drop in temperature with *r*. In other words the temperature distribution has a peak. Lin and Papaloizou (1985) have shown that thermal instabilities can produce a peaked temperature distribution in the solar disk. Perhaps thermal instabilities affected Uranus's disk in the same way.

Now $E_P$ values can be associated with the rest of Uranus's satellites by using the two close components of the η ring as a key. There should be a pair of rings or satellites with orbital radii larger than the radius of the peak in the temperature distribution that are associated with the same temperature and therefore $E_P$'s as the two close components of the η ring. The satellites 1986 U10 (Karkoschka 1999) and Belinda are found to be such a pair. Using them as the starting point, $E_P$'s are matched up with orbital radii to arrive at the rest of the data in Table 3. Fig. 4 is a graph of this data. Notice that the satellites Ophelia and Bianca presently orbit on either side, but close to the position of the maximum. These satellites have the same $E_P$ value associated with their resonances. Also, the closely spaced lines at 4822.8 and 4841.3 cm$^{-1}$ are associated with the closely spaced components of the η ring on one side of the maximum and with the satellites 1986 U10 and Belinda on the other side. The curves in Fig. 4 are fits to the data on each side,

$$E_P = 30{,}090/(r_0-r)^{0.194} \qquad r < 59{,}000 \text{ km} \qquad (8)$$

$$E_P = (5850/(r-r_0)^p + B_0) \qquad r > 59{,}000 \text{ km}, \qquad (9)$$

where $p = 0.196$ (interestingly, the same as for Saturn), $r_0 = 59{,}000$ km and $B_0 = 3970$ cm$^{-1}$. Equations like Eq. (9) will be found for Jupiter and Neptune in § 5. In preliminary fits, the $B_0$ on the right side of Eq. (9) and the $B_0$'s in the equations for Jupiter and Neptune are found to be nearly the same. Therefore they are finally constrained to be the same and found to be 3970 cm$^{-1}$. This is larger than the value 637 cm$^{-1}$ for the Sun and Saturn, indicating the possibility that the molecular association process for Uranus, Jupiter, and Neptune involves atoms or molecules that are different than for the Sun and Saturn.

Note the indices ($i$ = 1-14) in Table 3. Satellites of Uranus, Jupiter, and Neptune that are associated with the same $E_P$ value are assigned the same index in § 5 and § 6 below.

## 5. The Satellite System's of Jupiter and Neptune

Table 4 contains the data that relate $E_P$'s in the $H_2$ spectrum to the average orbital radii for the satellites of Jupiter and Neptune. This table does not include Jupiter's rings and Neptune's wide rings Lassell and Galle. These will be considered in § 9. Again, the indices $i$'s in Table 4 follow the scheme first mentioned for Uranus in § 4. The Jupiter data is generated by first noting the orbital radii of the satellites Metis and Adrastea are close, reminiscent of the close pairing of the two components of Uranus's η ring and the close pairing of Uranus's satellites 1986 U10 and Belinda. For this reason the orbital radii of Metis and Adrastea are matched with $E_P$ = 4822.8 and 4841.3 cm$^{-1}$. With this as a starting point all the successive satellites are matched with successive $E_P$'s. The equation that fits the Jupiter data is

$$E_P = (7010/(r-r_0)^p + B_0), \qquad (10)$$

where $p$ = 0.196 (the same as for Saturn and Uranus), $r_0$ = 84,200 km, $B_0$ = 3970 cm$^{-1}$ (the same as for Uranus) and $r$ is in kilometers from Jupiter's center. None of Jupiter's satellites or rings have orbital radii less than $r_0$ = 84,200 km, in contrast to Uranus where two satellites and many rings have radii less than $r_0$. A possible reason for this will be discussed in § 10.

The close-pair key used for Uranus and Jupiter can also be used for Neptune. Here the close pair is the satellite Galatea and ring Adams which leads to $E_P$'s being matched with average orbital radii of Neptune's satellites and rings as seen in Table 4. The equation that fits Neptune's data is

$$E_P = (5870/(r-r_0)^p + B_0), \qquad (11)$$

where $p$ = 0.196 (the same as for Saturn, Uranus and Jupiter!), $r_0$ = 46,300 km, $B_0$ = 3970 cm$^{-1}$ (the same as for Uranus and Jupiter!) and $r$ is in km from the Neptune's center. Plots of the Jupiter and Neptune data will be considered in the next section.

## 6. Radial Scaling

It is particularly useful to scale the radial coordinate $r$ according to

$$r_S = |r - r_0| / R_{scale} \qquad (12)$$

where

$$R_{scale} = (C/774)^{1/p}. \qquad (13)$$

In these formulas $p = 0.196$, $C$ is the constant in the numerator of each planet's effective collision energy function in Eqs. (7), (9), (10), and (11), $r_S$ is the scaled coordinate and $R_{scale}$ is the scaling constant. The constant $R_{scale}$ scales $|r - r_0|$ such that $r_S = 1$ when the effective collision energy equals 774 cm$^{-1}$, the value at 1 AU from the Sun (see Eq. (6)). The Sun's effective collision energy function is therefore used as a basis for the scaling. The formula for the effective collision energy function, $E_C{'}$, appropriate for all four giant planets becomes

$$E_C{'} = 774/r_S^{0.196} \quad cm^{-1}. \qquad (14)$$

This equation is obtained by using $|r - r_0| = r_S R_{scale}$ in the $E_C{'}$ term of any of the Eqs. (7), (9), (10) or (11).

Table 5 has the parameters, including scaling constants $R_{scale}$, for the giant planets. To discover the usefulness of radial scaling, the orbital radius $R$ of each of the satellites is substituted for $r$ in Eq. (12) and each of the $r_S$ values that is calculated is called a scaled orbital radius and designated $R_s$. This allows for consideration of graphs of effective collision energies ($E_C{'} = E_P - B_0$) vs. $R_s$ rather than $E_P$ vs. $R$. In doing so it is found that all the graphs for the giant planet systems are essentially the same, an unanticipated and wonderful result. For instance, consider Fig. 5, a graph of the effective collision energies for Uranus, Jupiter, and Neptune ($E_C{'} = E_P - 3970$ cm$^{-1}$). For this figure the $E_P$'s and scaled radii are taken from Table 6. Fig. 5 shows the points for the different planets corresponding to the same $E_C{'}$ value are in close agreement. Two adjacent points belonging to Jupiter's Amalthea and Thebe that don't fit the trend of the other points are indicated with arrows pointing in the direction of each one's apparent shift toward the other. In Fig. 5 there is one Uranus point in the upper left corner belonging to Bianca that is noticeable to the left of the fitted curve. Bianca is close to the peak in the collision energy function where we do not expect the fitted curve to follow the actual function because the actual function cannot diverge at r = $r_0$.

The indices ($i$ = 1-14) in Fig. 5 and Table 6 follow the same scheme as in Table 3 for Uranus and Table 4 for Jupiter and Neptune. To find the name of a satellite and the $H_2$ transition associated with a scaled radius in Fig. 5 or Table 6, use its index and refer back to Table 3 or 4.

Fig. 6 includes data plotted in Fig. 5 for Uranus, Jupiter, and Neptune plus the effective collision energies for Saturn ($E_C' = E_P - 637$ cm$^{-1}$). Saturn's $E_P$'s and scaled radii are taken from Table 1. Excellent alignment of the Saturn data with the other data is seen in the figure; each set of data being characterized by the same value of p = 0.196. The points for Amalthea and Thebe again have arrows in the direction of each one's apparent shift. This graph further indicates the existence of a common structure in the planets' effective collision energy functions, which is probably due to a common structure that existed in the temperature profiles of their primordial disks. The high degree of overlap among these graphs seems unlikely without there being validity in the assumed relationships between orbital radii and the spectra of atomic and molecular hydrogen. This is especially true in light of the fact that even though the profile for Saturn's protodisk was determined utilizing hydrogen energy levels and the profiles for the other giant planets' protodisks were determined utilizing the $H_2$ spectrum, the profiles are still the same.

The constant 3970 cm$^{-1}$ was determined by constraining it to have the same value in each of the fits to the Uranus, Jupiter, and Neptune data. In preliminary fits, the best-fit value for the constant was approximately 4000 cm$^{-1}$. The final value was taken to be 3970 cm$^{-1}$ because this caused an overlap of the Uranus, Jupiter, and Neptune data with the Saturn data in Fig. 6. Readjusting the parameter by this amount had very little effect on the quality of the fits.

## 7. Saturn's Rings and Binding Energies in $H_2^+$

In the present model the EINSER resonances for Saturn's D ring all correspond to molecules being formed in the same state. An interesting possibility is that the resonances for Saturn's other main rings all involve the same $E_P$'s as does the D ring ($n_f = 8$ in Eq. (5)) but the vibration-rotation state of the formed molecule, and therefore $E_B$, is different for each of the rings. If this is true, a graph of binding energies versus the corresponding radii of inner edges may be a smoothly varying curve. Furthermore, if the model is correct and if the correct match between binding energies and inner edge radii is found, it should be possible to fit this data with an equation that is related to Eq. (7), the equation already determined for Saturn. A particular sequence of binding energies in the $H_2^+$ molecular ion does in fact produce a smoothly varying curve that is related to Eq. (7) as is seen in the following discussion.

Moss (1993) has calculated the binding energies of the vibration-rotation levels of the $H_2^+$ molecular ion. Each state is characterized by quantum numbers ($v, J$). Careful consideration of Moss's results reveals one sequence of states with binding energies that are correlated to the inner edges of Saturn's main rings. The sequence includes the (15,0) state and the eight neighboring $v = 14$ states with binding energies greater than that of the (15,0) state. Table 7 includes the sequence with each binding energy paired with the inner edge radius $R_{ie}$ of a ring. The first five $R_{ie}$ values correspond to the inner edges of the D, C, B, A, and G rings. The sixth one corresponds to the inner edge of the largest peak in the E ring detected by images of ring brightness taken by Bauer et al. (1997), which in the present model is actually a ring superimposed on the E ring. Fig. 7 is a graph of data in Table 7 including the first six pairings (filled circles) and three predictions discussed below (open circles). The graph is rather smooth. More importantly though, it can be fitted by an equation deduced from Eq. (7). To find this equation, Eq. (7) is written in the following way for the specific case of the inner edge of the D ring, i.e. $E_P = 1714.7$ cm$^{-1}$,

$$1714.7 = (8032/(R_{ieD} - r_0)^p + 848.8/1.332) \text{ cm}^{-1}, \tag{15}$$

where $r_0 = 37,800$ km, $p = 0.196$ and the constant 637 cm$^{-1}$ in Eq. (7) has been replaced by its equivalent value 848.8 cm$^{-1}$ divided by 1.332. The value 848.8 cm$^{-1}$ is the binding energy that is paired with the inner edge of the D ring in Table 7 and $R_{ieD}$ is the radius of the inner edge of the D ring. Now under the assumption that Eq. (15) is a specific case of a more general formula relating binding energies to inner edge radii $R_{ie}$, the constant 848.8 cm$^{-1}$ is replaced by $E_B$ and $R_{ieD}$ by $R_{ie}$. Solving for $E_B$ yields

$$E_B = (2284 - 10,700/(R_{ie} - r_0)^p). \tag{16}$$

The curve in Fig. 7 is a graph of Eq. (16). The fit to the data is very good indicating that both Saturn's satellite data in Table 1 and its ring data in Table 7 can be fitted by the single formula obtained by generalizing Eq. (15)

$$E_P(n_f, n_i) = 8032/(r - r_0)^p + E_B(v, J)/1.332. \tag{17}$$

That both sets of data are fitted with one equation indicates the uniqueness of the sequence of $H_2^+$ binding energies in Table 7 and serves as a check of the model. No other sequence in $H_2^+$ was found to be related to the radius data. These results indicate the EINSER resonances associated with each main ring of Saturn are all created with the same $E_P$'s but each with a different $E_B$. They also indicate that in the case of Saturn and presumably the Sun, because $B_0$ also equals 637 cm$^{-1}$ for the Sun, that H and H$^+$ merge to form $H_2^+$ in the EINSER resonance process.

The last three $R_{ie}$'s in Table 7 and the open circles in Fig. 7 are predictions made using Eq. (16) with $E_B = E_B(v=14; J=0, 1, \text{and } 2)$. They are 2.82x10$^5$ km, 3.37x10$^5$ km, and 3.72x10$^5$ km. Interestingly, these predictions have values close to the positions of the inner edges of the next three largest peaks detected in E ring brightness. Reading from a graph given by Bauer et al. (1997) these peaks have inner edge radii of 2.9x10$^5$ km, 3.3x10$^5$ km, and 3.7x10$^5$ km.

Consider once again the states listed in Table 7. States with quantum number $v$ = 14 and 15 have binding energies which are intermediate among all the binding energies of $H_2^+$ states; the least bound state has quantum number $v$ = 19 (Moss 1993). A possible explanation for why resonance occurs for the formation of these intermediate $v$ = 14 and 15 states is that cross sections are highest in this range of states. A precedent for intermediate states having the highest cross sections is seen in the theoretical work of Gianturco and Giorgi (1997) involving radiative association rates for Li and $Li^+$ colliding with H and $H^+$.

## 8. The Collision Energy Function

The collision energy is dependent on the temperature and therefore the average kinetic energy of the particles in the disk. Because the temperature is radially dependent the collision energy depends on $r$. But Eq. (17) indicates the possibility of the collision energy function also depending on the energy of the photon that stimulates the radiative association process. To show this both sides of Eq. (17) are multiplied by 1.332 and rearrangement gives

$$E_P = (10{,}700/(r - r_0)^p - 0.332\, E_P) + E_B. \qquad (18)$$

From Eqs. (3) and (18) the actual (not effective) collision energy function is

$$E_C = 10{,}700/(r - r_0)^p - 0.332\, E_P. \qquad (19)$$

This equation indicates how the collision energy depends not only on the radially dependent temperature but also how it apparently depends on $E_P$. J. C. Lombardi, Jr. (private communication) has pointed out that the dependence of $E_C$ on $E_P$ may actually be an artifact, possibly produced because the EINSER resonance for each ring is produced at a different time and the collision energy function changes slightly over the time when the various resonances are occurring. On the other hand the possibility of $E_C$ depending directly on $E_P$ and even on $E_B$, is very interesting and warrants further investigation.

Eq. (19) gives a negative and therefore impossible result for $E_C$ if the second term on the right side has an absolute value that is larger than the first term. For the range of $E_P$'s used in this investigation this problem does not arise.

## 9. The Wide Rings of the Giant Planets

The wide rings of Jupiter and Neptune apparently resulted from EINSER resonances involving the same series of $E_P$'s in the spectrum of atomic hydrogen that generated the main rings of Saturn. This is shown by scaling the radii of the inner and outer edges of the rings of Saturn, Jupiter, and Neptune using Eq. (12) and the parameters in Table 5. The observed inner and outer radii are designated by $R_{ie}$ and $R_{oe}$ and the scaled radii are designated by $R_{Sie}$ and $R_{Soe}$. The results are listed in Table 8 and used to make Fig. 8 where scaled radii for rings of each planet are plotted along a separate line. A beauty of radial scaling is rings around different planets that are associated with the same $E_P$'s and $E_C$''s have close to the same radially scaled coordinates for their inner and outer radii. In light of this there are a few striking features of the graphs in Fig. 8. Firstly, the B, C and D rings of Saturn span nearly the same region as the halo around Jupiter, with the ring Lassell of Neptune also centered in that region. Secondly, the position of the A ring of Saturn corresponds closely to the main ring of Jupiter. Finally the E ring of Saturn and the gossamer ring of Jupiter both span approximately the same distance in units of scaled radii (1.1 units). However they are shifted with respect to one another, as if one or both have migrated since their creation. Listed in Table 8 but not shown in Fig. 8 is Galle, the other wide ring of Neptune. This ring was formed in the region where $r$ is less than $r_0$ (for Neptune $r_0 = 46,300$ km) so it can't be scaled in the same way as the other wide rings. However, Galle may be a "reflection" of Lassell on the other side of the maximum in the collision energy function in the same way Uranus's rings are "reflections" of satellites that exist beyond $r_0$ in Uranus. The similarity in the patterns illustrated in Fig. 8 is an indication that a common mechanism initiated the evolution of these rings.

## 10. Further Discussion

As mentioned at the end of § 2, other schemes matching $E_P$'s with orbital radii in the solar system have been investigated in addition to the one given in Table 1. For example, shifting each $E_P$ by one position so that the $E_P$ associated originally with Venus becomes the $E_P$ for Earth etc. generates another set of data. The resulting scheme is not nearly as suitable as the original one for a few reasons. In the second scheme: (1) The calculated inner edge radius of the asteroid belt is between Vesta and Juno not between Mars and Vesta as it should be. (2) The calculated inner edge of the Kuiper Belt is 55 AU well beyond the observed value of 42 AU. (3) A graph of $E_P$ vs. orbital radii is not a smooth curve as in Fig. 1. The data points for the asteroids Vesta, Eunomia, Juno, Bamberga, Ceres and Pallas fall well below a curve that is a fit to the planets. (4) There are two close orbital radii associated with Neptune and not with Uranus as in the original scheme. In the original scheme the proximity of the radii helps to explain the tilt of the Uranus's axis of rotation (§ 2.5). Shifting $E_P$'s one position in the other direction with respect to the orbital radii produces similar undesirable results. Shifts of more than one position produces schemes that match the solar system even more poorly.

The spectrum of atomic hydrogen has been shown to be key to the creation of Saturn's satellite system. There are still three $E_P$'s left to consider that correspond to radii within the system, and these radii all fall within Saturn's rings: $E_P(7,13) = 1590.2$ cm$^{-1}$, $E_P(7,14) = 1679.7$ cm$^{-1}$, and $E_P(6,9) = 1693.5$ cm$^{-1}$. Using these $E_P$'s and Eq. (7) the radii 9.06x10$^4$ km, 7.12x10$^4$ km, and 6.90x10$^4$ km are calculated. The first of these is within the C ring where there are so many ringlets it is impossible to determine if the $E_P(7,13)$ resonance has had an effect there. The other two radii are within the diffuse D ring. The structure of the D ring is characterized mainly by two narrow ringlets with radii of 7.17x10$^4$ km and 6.76x10$^4$ km and a fainter and broader feature at 7.31x10$^4$ km (Showalter 1996). The second and third calculated radii are close to the observed radii of the narrow ringlets, thus indicating that these particular features are due to the $E_P(7,14)$ and $E_P(6,9)$ EINSER resonances.

Equation (3) indicates EINSER activity will occur with $E_P$ nearly equal to $E_B$ where $E_C$ and therefore the temperature of the disk is nearly zero. With this in mind Fig. 1 indicates the temperature of the Sun's disk is near zero for $r$ approximately equal to 50 AU. This is the value of $r$ where $E_P = E_B =$ 848.8 cm$^{-1}$ (the binding energy of the $E_B(15,0)$ state in $H_2^+$). The value 50 AU can be considered a measure of the radius of the disk and therefore of the solar system. Jewitt et al. (1998) carried out a survey to discover Kuiper Belt Objects. They report "We are uncomfortable with the notion that the Kuiper Belt might have an edge near 50 AU (what physical process could be responsible?) but our data nevertheless suggest this as a possibility." It seems that the EINSER model supplies an explanation for the 50 AU limit.

The principles of conservation of momentum and energy can be used to show in a molecular association process that occurs without stimulation, on average the collision energy is equal to the average kinetic energy that each particle has before they merge. In a gas the average kinetic energy and therefore the collision energy is 3/2kT. The collision energy function for Saturn given by Eq. (19) involves the energy of the stimulating photon and it is therefore more complicated than would be expected for an association process without stimulation. Assuming for Saturn that just the first term on the right side of Eq. (19) is approximately equal to 3/2kT, the following approximate relationship for the temperature in Saturn's disk as a function of the scaled radial coordinate is found,

$$T \sim 989/r_S^{0.196} \quad K. \qquad (20)$$

This relationship was derived for Saturn and is used to estimate temperatures corresponding to various points in Fig. 6, the graph that gives the composite effective collision energy function for the giant planets. The three points that are labeled with temperature values, define the beginning and end positions of dips in the function. The known $r_S$ values for each of these points are used in Eq. (20) to yield the approximate temperatures $T_A = 1370$ K, $T_B = 1120$ K, and $T_C = 840$ K. Cameron (1985) has made calculations of temperature distributions in the solar nebula for various conditions in the nebula. Each distribution has a single dip in the radial direction starting at a temperature of about 1600 K and extending down below 1000 K. The dip is related to the temperature at which solids condense with iron condensing at temperatures below approximately 1300 K. The temperatures indicated in Fig. 6 are in the range of the dips in the Cameron calculations and therefore may be related to the same effects he used to determine structure in the solar nebula's temperature distribution. Possibly thermal instabilities similar to those in the solar disk (Lin and Papaloizou 1985) have also affected the structure.

The radius parameter $r_0$ is varied in each of the fits for the giant planets. Table 5 contains ratios $r_0/R_P$, where $R_P$ designates the equatorial radius of a planet. For Uranus and Neptune the ratios are 2.3 and 1.9 and for Saturn and Jupiter they are somewhat smaller at 0.63 and 1.2 respectively. For the Sun the ratio is zero since $r_0$ is zero. The ratios for the giant planets help to explain why Uranus and Neptune have rings inside the radius $r_0$ while Saturn and Jupiter do not. Fig. 4 shows the distribution of the effective collision energy function for Uranus is peaked in the vicinity where the orbital radius is $r_0$. Uranus's rings are left of the peak. The ratio $r_0/R_P$ is large enough for the left side of the distribution to exist beyond the planet's surface and rings are possible there. A similar situation exists for Neptune and its ring Galle is inside $r_0$. In the cases of Saturn and Jupiter the ratios $r_0/R_P$ are smaller and there is little if any room for the left side of the distribution to exist outside the planet. Therefore there are no rings inside the radius $r_0$ for Saturn and Jupiter. Also, these ratios may someday help to explain why for Saturn and the Sun, where the ratios are less than 1, the $E_P$'s that fit the data are in the spectrum of atomic hydrogen and for Uranus, Neptune, and Jupiter, where the ratios are larger than 1, the $E_P$'s are mainly in the spectrum of molecular hydrogen. Furthermore, the relatively large value of $r_0/R_P$ for Uranus and the resulting unique ring and satellite system of the planet may be related to Uranus's creation by the collision and merger of two protoplanets (§ 2.5) before EINSER activity was initiated in the planet's disk.

## 11. Remaining Questions

Besides the questions dealing with the ratio $r_0/R_P$ there are others that require investigation:

1. What molecule is formed in the EINSER activity associated with the planets Uranus, Jupiter, and Neptune (§ 4 and § 5)?
2. What mechanism has created the structure in the effective collision energy function seen in Fig. 6?
3. Why is $p$ equal to 0.108 for the solar system and a different value 0.196 for all the giant planets?
4. Why does the solar system analysis require the $-0.13/r^2$ term and the analyses for the planetary systems do not (§ 2.1)?
5. Are the many ringlets in Saturn's rings due to the effect of EINSER resonance?
6. Would observations near 26 AU from the Sun reveal a remnant of the belt in the Uranus-Neptune zone predicted by the present analysis (§ 2.4)? Holman's simulations (1997) have identified this as a region of stability for the age of the solar system.
7. Can EINSER radiation be detected from stellar disks?
8. Can EINSER radiation be created in the laboratory?

## 12. Summary

The present model explains many of the characteristics of the solar system including the orbital radii of the planets and its satellites, the tilt of Uranus's axis, the positions of the asteroid and Kuiper belts, the source of the scattered Kuiper belt objects, the positions of Saturn's main rings and the rings of Uranus, Jupiter, and Neptune. It also shows that a commonality exists in the structures of the solar system and the planetary systems that can be attributed to the common process that initiated their evolution. The almost identical structure in the effective collision energy functions for the giant planets seen in Fig. 6 is most illustrative of this.

The results of the present investigation imply there are many stars with similar planetary systems. Indeed, any protostar that has similar characteristics to the protosun would generate EINSER resonances in its primordial disk, thus initiating an evolution like that in the solar system.

Acknowledgments

I especially want to thank my son James C. Lombardi Jr. for helpful conversations and for encouragement to continue work on this project after I had put it aside for ten years. I acknowledge James Hilton for kindly providing data concerning the largest objects in the asteroid belt and Phillip Stancil for his unpublished calculations of binding energies in $H_2^+$ and for helpful conversations concerning the radiative association of molecules. I am very appreciative of Paul Martini for supplying me with the file used to make Fig. 3. If not for the Martini, Sellgren, DePoy graph I might not have seen the relationship between the $H_2$ spectrum and the average orbital radii of the satellites and rings around Uranus, Jupiter, and Neptune.

TABLE 1.

Photon energies and orbital radii in the solar and Saturnian systems.

| $(n_f,n_i)$ | $E_P(n_f,n_i)$ (cm$^{-1}$) | Solar System | $R$(AU)[a] | Saturn's System | $R$(km)[i] | $R_S$[j] |
|---|---|---|---|---|---|---|
| (8,∞) | 1714.7 | I. E.[b] not observed | (0.047)[c] | I. E. D ring | 66,000[k] | 0.185 |
| (8,22) | 1487.9 | Mercury's Region | 0.387 | Pan | 133,583 | 0.627 |
| (7,12) | 1477.5 | Mercury's Region | 0.387 | Atlas | 137,640 | 0.653 |
| (8,21) | 1465.8 | Mercury's Region | 0.387 | Prometheus | 139,350 | 0.665 |
| (8,20) | 1440.3 | Venus | 0.723 | Pandora | 141,700 | 0.680 |
| (8,19) | 1410.7 | Earth | 1.000 | Epimetheus and Janus | 151,500 | 0.744 |
| (8,18) | 1376.0 | Mars | 1.524 | Mimas | 185,600 | 0.967 |
| (9,∞) | 1354.8 | I. E. Asteroid Belt | 2.0[d] | I. E. E-Ring | ~180,000[l] | |
| (5,6) | 1341.2 | Vesta | 2.361[e] | Enceladus | 238,100 | 1.310 |
| (8,17) | 1334.9 | E, J and B[f] | 2.67[e] | Tethys | 294,700 | 1.681 |
| (6,8) | 1333.6 | Ceres | 2.768[e] | Telesto | 294,700 | 1.681 |
| (7,11) | 1332.6 | Pallas | 2.774[e] | Calypso | 294,700 | 1.681 |
| (8,16) | 1286.0 | Jupiter | 5.203 | Dione and Helene | 377,400 | 2.223 |
| (8,15) | 1226.9 | Saturn | 9.537 | Rhea | 527,100 | 3.202 |
| (8,14) | 1154.8 | Uranus | 19.19 | Titan | 1,221,900 | 7.749 |
| (7,10) | 1142.2 | Uranus | 19.19 | Hyperion | 1,464,100 | 9.445 |
| (10,∞) | 1097.4 | I. E. U-N zone | (25.6)[g] | I. E. not observed | --- | |
| (8,13) | 1065.3 | Neptune | 30.07 | Iapetus | 3,560,800 | 23.06 |
| (8,12) | 952.58 | Pluto | 39.48 | | | |
| (11,∞) | 906.90 | I. E. Kuiper Belt | 42[h] | | | |

[a] Planetary Mean Orbits J2000 (2003). In JPL's *Solar System Dynamics* except where indicated otherwise. http://ssd.jpl.nasa.gov/elem_planets.html
[b] I. E. stands for inner edge.
[c] Predicted inner edge of a belt.
[d] Clark (1999).
[e] Hilton J. L. (2003) In *U. S. Naval Observatory Ephemerides of the Largest Asteroids*, http://aa.usno.navy.mil/ephemerides/asteriod//astr_alm/asteroid_ephemeredes.html.
[f] E, J and B stands for Eunomia, Juno and Bamberga.
[g] Predicted inner edge of a belt in the Uranus-Neptune zone
[h] Jewitt (1999).
[h] Jacobsen (1996) and Jacobson R. A. (2003) In SAT077 – JPL satellite ephemeris. http://ssd.jpl.nasa.gov/ except where otherwise indicated.
[j] The scaled radii $R_S$ for Saturn are discussed in § 6.
[k] Showalter (1996).
[l] de Pater et al. (1996); Bauer et al. (1997).

TABLE 2.

Solar system belts.

| Series Limit | Calculated Inner Radius (AU) | Observed Inner Radius (AU) | Calculated Outer Radius (AU)[a] | Observed Outer Radius (AU) | Belt Name |
|---|---|---|---|---|---|
| $E_P(8, \infty)$ | 0.047 | ---- | 0.075 | ---- | not observed |
| $E_P(9, \infty)$ | 2.0 | 2.0[b] | 4.0 | 4.0[b] | asteroid belt |
| $E_P(10, \infty)$ | 25.6 | ---- | 31.3 | ---- | belt in U-N zone |
| $E_P(11, \infty)$ | 43.6 | 42.[c] | 47.6 | 47.[c] | Kuiper belt |

[a] Calculated outer radii correspond to $E_P(n_f, n_i = 45)$.
[b] Clark (1999)
[c] Jewitt (1999)

TABLE 3.

Orbital radii of rings and satellites of Uranus with $E_P$'s for $H_2$ transitions.

| i | $E_P$[a] (cm$^{-1}$) | $H_2$ Transition[a] | Satellite or ring | $R < 59000$ km $R$ (km) [b,c] | Satellite | $R > 59000$ km $R$ (km) [b] |
|---|---|---|---|---|---|---|
| 1 | 5397.2 | (2,1) S(7) | Ophelia | 53,763 | Bianca | 59,166 |
| 2 | 5285.6 | (1,0) S(4) | ring epsilon | 51,149 | Cressida | 61,767 |
| 3 | 5146.8[d] | (9,7) S(1) | ring λ | 50,024 | Desdemona | 62,658 |
| 3 | 5141.8[d] | (2,1) S(5) | ring λ | 50,024 | Desdemona | 62,658 |
| 4 | 5108.4 | (1,0) S(3) | Cordelia | 49,752 | Juliet | 64,358 |
| 5 | 4989.8 | (2,1) S(4) | ring δ | 48,300 | Portia | 66,097 |
| 6 | 4917.0 | (1,0) S(2) | ring γ | 47,627 | Rosalind | 69,927 |
| 7 | 4841.3 | (3,2) S(5) | ring η$_2$ | 47,240 | Belinda | 75,255 |
| 8 | 4822.8 | (2,1) S(3) | ring η$_1$ | 47,176 | 1986 U 10 | 76,416[e] |
| 9 | 4712.9 | (1,0) S(1) | ring β | 45,661 | Puck | 86,004 |
| 10 | 4642.1 | (2,1) S(2) | ring α | 44,718 | Miranda | 129,870 |
| 11 | 4542.6 | (3,2) S(3) | ring 4 | 42,571 | Ariel | 190,950 |
| 12 | 4497.8 | (1,0) S(0) | ring 5 | 42,235 | Umbriel | 266,000 |
| 13 | 4449.0 | (2,1) S(1) | ring 6 | 41,837 | Titania | 436,300 |
| 14 | 4372.4 | (3,2) S(2) | ring U2R | 38,250 | Oberon | 583,520 |

[a] Black and van Dishoeck (1987).
[b] Orbital radii are from Jacobson (1998), and Laskar and Jacobson (1987) except as noted.
[c] Ring radii are from French et al. (1991).
[d] The close $E_P$ values 5141.8 and 5146.8 cm$^{-1}$ are averaged in the analyses.
[e] Karkoschka (1999).

TABLE 4.

Orbital radii of satellites and rings of Jupiter and Neptune with $E_P$'s for $H_2$ transitions.

| $i$ | $E_P$[a] ($cm^{-1}$) | $H_2$ Transition[a] | Jupiter Satellite | $R(km)$[b] | Neptune Satellite or Ring | $R(km)$[c] |
|---|---|---|---|---|---|---|
| 2 | 5285.6 | (1,0) S(4) | | | Naiad | 48,227 |
| 3 | 5146.8[d] | (9,7) S(1) | | | Thalassa | 50,075 |
| 3 | 5141.8[d] | (2,1) S(5) | | | Thalassa | 50,075 |
| 4 | 5108.4 | (1,0) S(3) | | | Despina | 52,526 |
| 5 | 4989.8 | (2,1) S(4) | | | ring LeVerrier | 53,200 |
| 6 | 4917.0 | (1,0) S(2) | | | ring Arago | 57,600 |
| 7 | 4841.3 | (3,2) S(5) | Metis | 127,960 | Galatea[e] | 61,953 |
| 8 | 4822.8 | (2,1) S(3) | Adrastea | 128,980 | ring Adams | 62,933 |
| 9 | 4712.9 | (1,0) S(1) | Amalthea | 181,300 | Larissa | 73,548 |
| 10 | 4642.1 | (2,1) S(2) | Thebe | 221,900 | Proteus | 117,647 |
| 11 | 4542.6 | (3,2) S(3) | Io | 421,770 | | |
| 12 | 4497.8 | (1,0) S(0) | Europa | 671,080 | | |
| 13 | 4449.0 | (2,1) S(1) | Ganymede | 1,070,040 | | |
| 14 | 4372.4 | (3,2) S(2) | Callisto | 1,883,000 | | |

[a] Black and van Dishoeck (1987)
[b] Jacobson R. A. (2003) In JUP204 – JPL satellite ephemeris. http://ssd.jpl.nasa.gov/
[c] Satellite orbital radii from Owen et al. (1991). Ring radii from Porco et al. (1995).
[d] The close $E_P$ values 5141.8 and 5146.8 $cm^{-1}$ are averaged in the analyses.
[e] Galatea's orbital radius is the same as the radius of a very faint unnamed ring.

TABLE 5.

Model parameters and equatorial radii of planets.

| | $p$ | $C$ | $B_0$ | $r_0(km)$ | $R_{scale}(km)$ | $R_P(km)$[a] | $r_0/R_P$ |
|---|---|---|---|---|---|---|---|
| Saturn | 0.196 | 8032 | 637 | 37,800 | 152,800 | 60,268 | 0.627 |
| Uranus | 0.196 | 5850 | 3970 | 59,000 | 30,320 | 25,559 | 2.308 |
| Jupiter | 0.196 | 7010 | 3970 | 84,200 | 76,300 | 71,492 | 1.178 |
| Neptune | 0.196 | 5870 | 3970 | 46,300 | 30,850 | 24,764 | 1.870 |

[a] JPL's Horizons System (2003). http://ssd.jpl.nasa.gov/horizon.html

TABLE 6.

Photon energies, effective collision energies, and scaled radii
for the satellites of Uranus, Jupiter and Saturn.

| i | $E_P$(cm$^{-1}$) | $E_C'$(cm$^{-1}$) | Uranus[a] $R_S$ | Jupiter $R_S$ | Neptune $R_S$ |
|---|---|---|---|---|---|
| 1 | 5397.2 | 1427.2 | 0.00548 | | |
| 2 | 5285.6 | 1315.6 | 0.0913 | | 0.0625 |
| 3 | 5144.3 | 1174.3 | 0.121 | | 0.122 |
| 4 | 5108.4 | 1138.4 | 0.177 | | 0.202 |
| 5 | 4989.8 | 1019.8 | 0.234 | | 0.224 |
| 6 | 4917.0 | 947.0 | 0.360 | | 0.366 |
| 7 | 4841.3 | 871.3 | 0.536 | 0.574 | 0.507 |
| 8 | 4822.8 | 852.8 | 0.574 | 0.587 | 0.539 |
| 9 | 4712.9 | 742.9 | 0.891 | 1.273 | 0.883 |
| 10 | 4642.1 | 672.1 | 2.334 | 1.805 | 2.313 |
| 11 | 4542.6 | 572.6 | 4.361 | 4.422 | |
| 12 | 4497.8 | 527.8 | 6.820 | 7.689 | |
| 13 | 4449.0 | 479.0 | 12.43 | 12.92 | |
| 14 | 4372.4 | 402.4 | 17.27 | 23.58 | |

[a] For Uranus only the radii corresponding to $R > r_0$ are included.

TABLE 7.

Inner edge radii of Saturn's rings and binding
energies of states in $H_2^+$.

| v | J | $E_B(v, J)$(cm$^{-1}$)[a] | Ring | $R_{ie}$(km) |
|---|---|---|---|---|
| 15 | 0 | 848.76 | D | 66,000[b] |
| 14 | 7 | 909.32 | C | 74,510[c] |
| 14 | 6 | 1024.4 | B | 92,000[c] |
| 14 | 5 | 1126.9 | A | 122,170[c] |
| 14 | 4 | 1215.0 | G | 166,000[d] |
| 14 | 3 | 1287.4 | Peak in E | 217,000[e] |
| 14 | 2 | 1342.7 | Prediction | 282,000 |
| 14 | 1 | 1380.1 | Prediction | 337,000 |
| 14 | 0 | 1398.9 | Prediction | 372,000 |

[a]Moss (1993).
[b]Showalter (1996).
[c]Cuzzi et al. (1984).
[d]Showalter and Cuzzi (1993).
[e]Bauer et al. (1997).

TABLE 8.

Observed and scaled radii for rings of Saturn, Jupiter and Neptune.

| Saturn | | | | |
|---|---|---|---|---|
| Ring | $R_{ie}$(km) | $R_{oe}$(km) | $R_{Sie}$ | $R_{Soe}$ |
| D | 66000[a] | --- | 0.185 | --- |
| B | --- | 117,580[b] | --- | 0.522 |
| A | 122,170[b] | 136,780[b] | 0.552 | 0.648 |
| E | 180,000[c] | 360,000[c] | 0.931 | 2.109 |

| Jupiter | | | | |
|---|---|---|---|---|
| Ring | $R_{ie}$(km) | $R_{oe}$(km) | $R_{Sie}$ | $R_{Soe}$ |
| Halo | ~93,000[d] | ~122,000[d] | ~0.12 | ~0.50 |
| Main | 123,000[d] | 129,130[e] | 0.509 | 0.589 |
| gossamer | 129,130[e] | 210,000[f] | 0.589 | 1.65 |

| Neptune | | | | |
|---|---|---|---|---|
| Ring | $R_{ie}$(km) | $R_{oe}$(km) | $R_{Sie}$ | $R_{Soe}$ |
| Galle | 41,000[g] | 43,000[g] | --- | --- |
| Lassell | 53,200[g] | 57,200[g] | 0.224 | 0.353 |

[a]Showalter (1996).
[b]Cuzzi et al. (1984).
[c]de Pater et al. (1997).
[d]Burns et al. (1984).
[e]Showalter et al. (1987).
[f]Showalter et al. (1985).
[g]Porco et al. (1995).

Figure Captions

Figure 1. Photon energies $E_P$ vs. orbital radii $R$ in the solar system. The curve is a graph of equation (6). The calculated inner edge radii of three belts discussed in § 2.2 – § 2.4 are also indicated.

Figure 2. Photon energies $E_P$ vs. orbital radii $R$ in Saturn's system. The curve is a graph of equation (7). The inner edge radius of the D ring as well as the calculated inner and outer edge radii of the E ring are discussed in § 3.1 and indicated in this figure.

Figure 3. Model spectrum of $H_2$ (Draine and Bertoldi 1996) smoothed by Martini, Sellgren, and DePoy. (1999). Peaks corresponding to S transitions are matched with rings or satellites near Uranus showing a pattern of peak positions similar to the pattern of ring and satellite positions.

Figure 4. Photon energies $E_P$ vs. orbital radii $R$ in Uranus's system. The curves are graphs of equations (8) and (9). Points at directly opposite positions on either side of the peak correspond to satellites with the same value of $E_P$. The two components of the η ring and the pair, Belinda and 1986 U10 serve as the key to deciphering the relationship between $E_P$'s and the average orbital radii of Uranus, Jupiter and Neptune.

Figure 5. Effective collision energies $E_C'$ vs. scaled radii $R_S$ for Uranus, Jupiter and Neptune showing the overlap of effective collision energy functions. The positions of the points for Jupiter's Amalthea and Thebe indicate these planets may have migrated toward each other. The curve is a graph of equation (14). Bianca is a satellite near the peak in Uranus's effective collision energy function and therefore its point is not well fitted by the curve.

Figure 6. Composite effective collision energy function for the giant gas planets. The indicated approximate temperatures are calculated from equation (20). At least three dips in the distribution are identifiable, and they indicate a common structure in the planets' effective collision energy functions.

Figure 7. Binding energies of the six $H_2^+$ states are graphed against inner edge radii of Saturn's rings (filled circles). The curve is a graph of equation (16). The open circles are predictions made using equation (16) and the remaining three binding energies in the sequence discussed in § 7.

Figure 8. Wide Rings of the Giant Planets. Scaled radii of inner and outer edges of the wide rings are plotted for Saturn, Jupiter and Neptune. Rings associated with the same $E_P$ and $E_C'$ values fall in approximately the same region of space along the scaled radial direction.

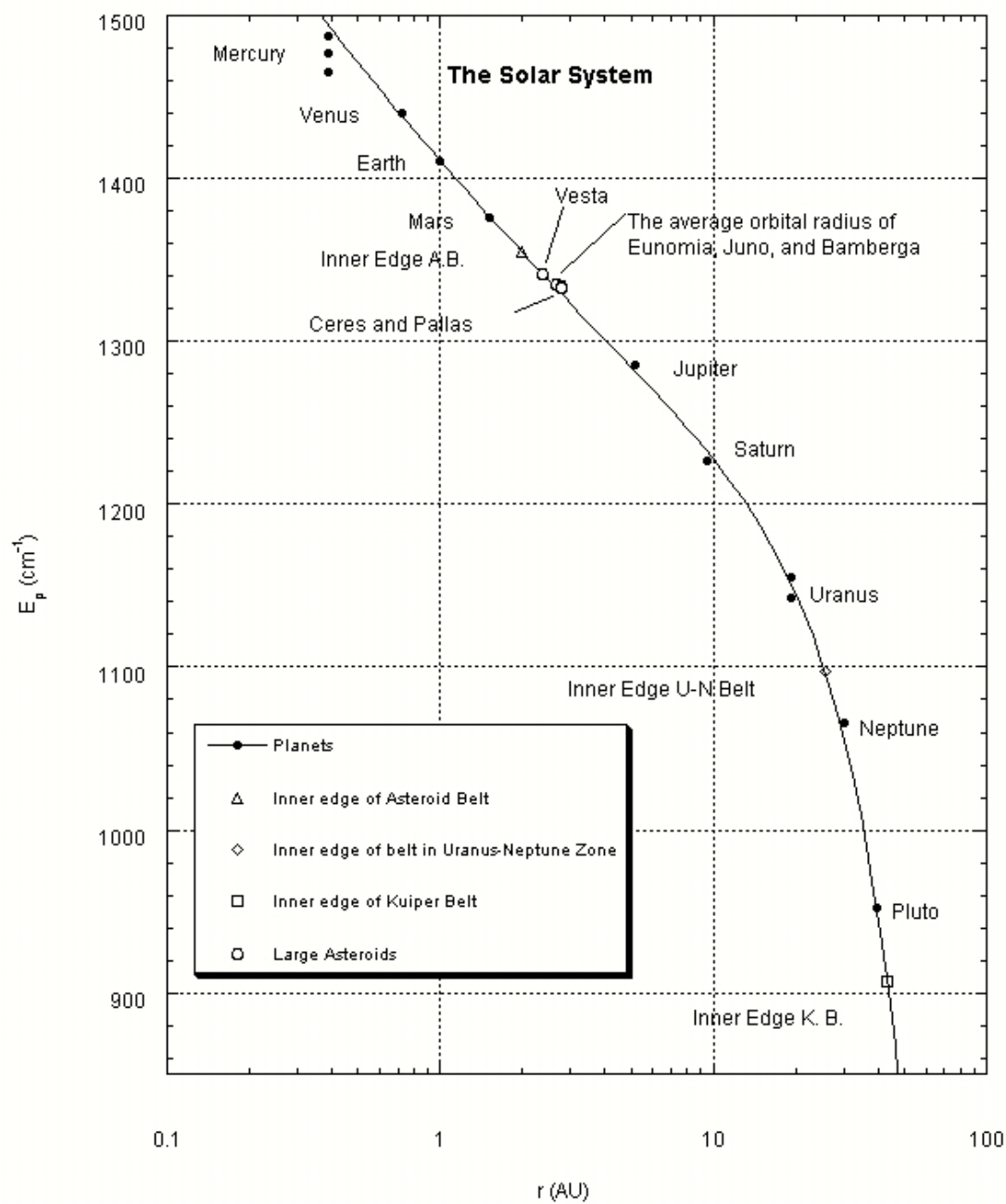

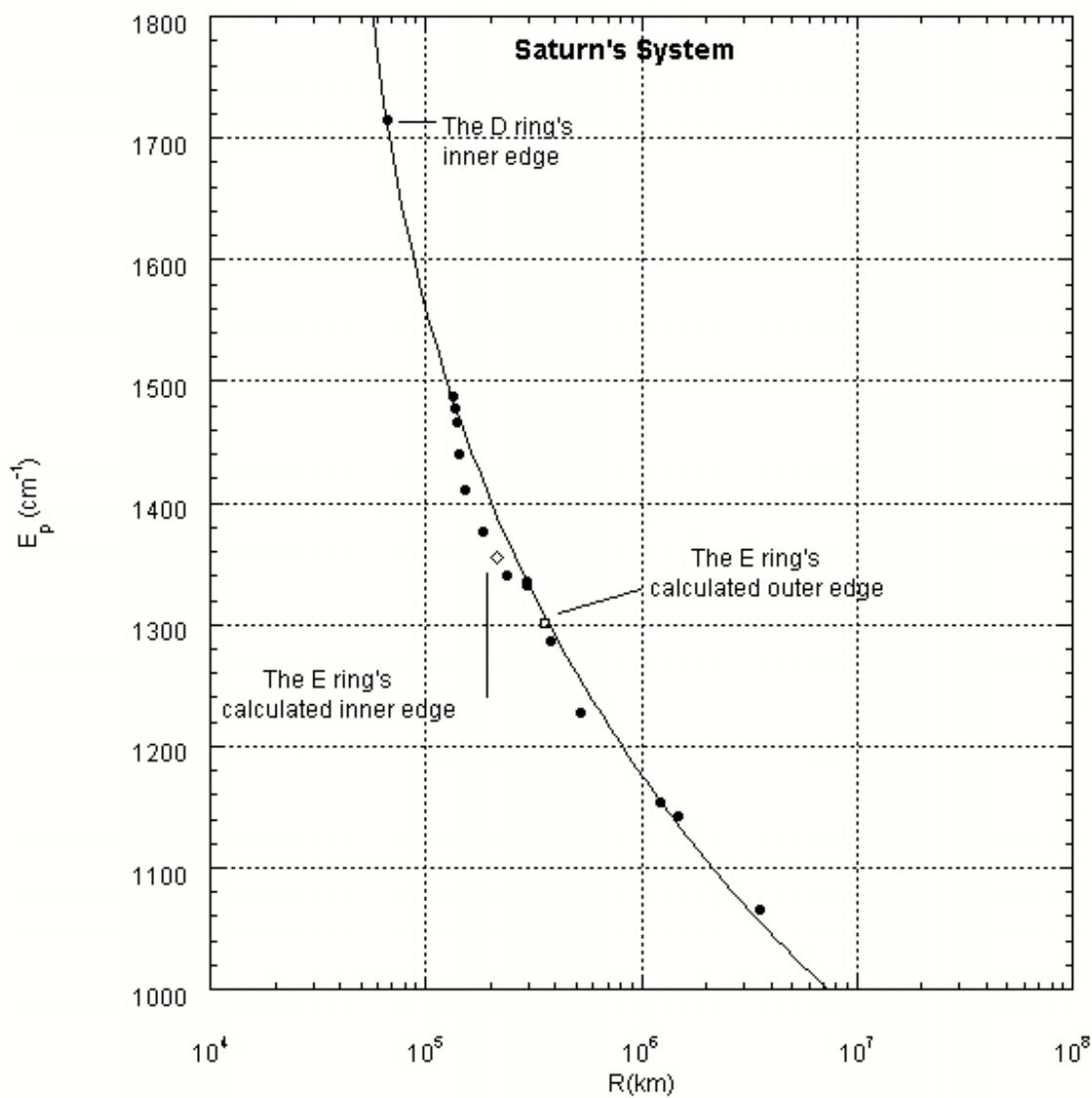

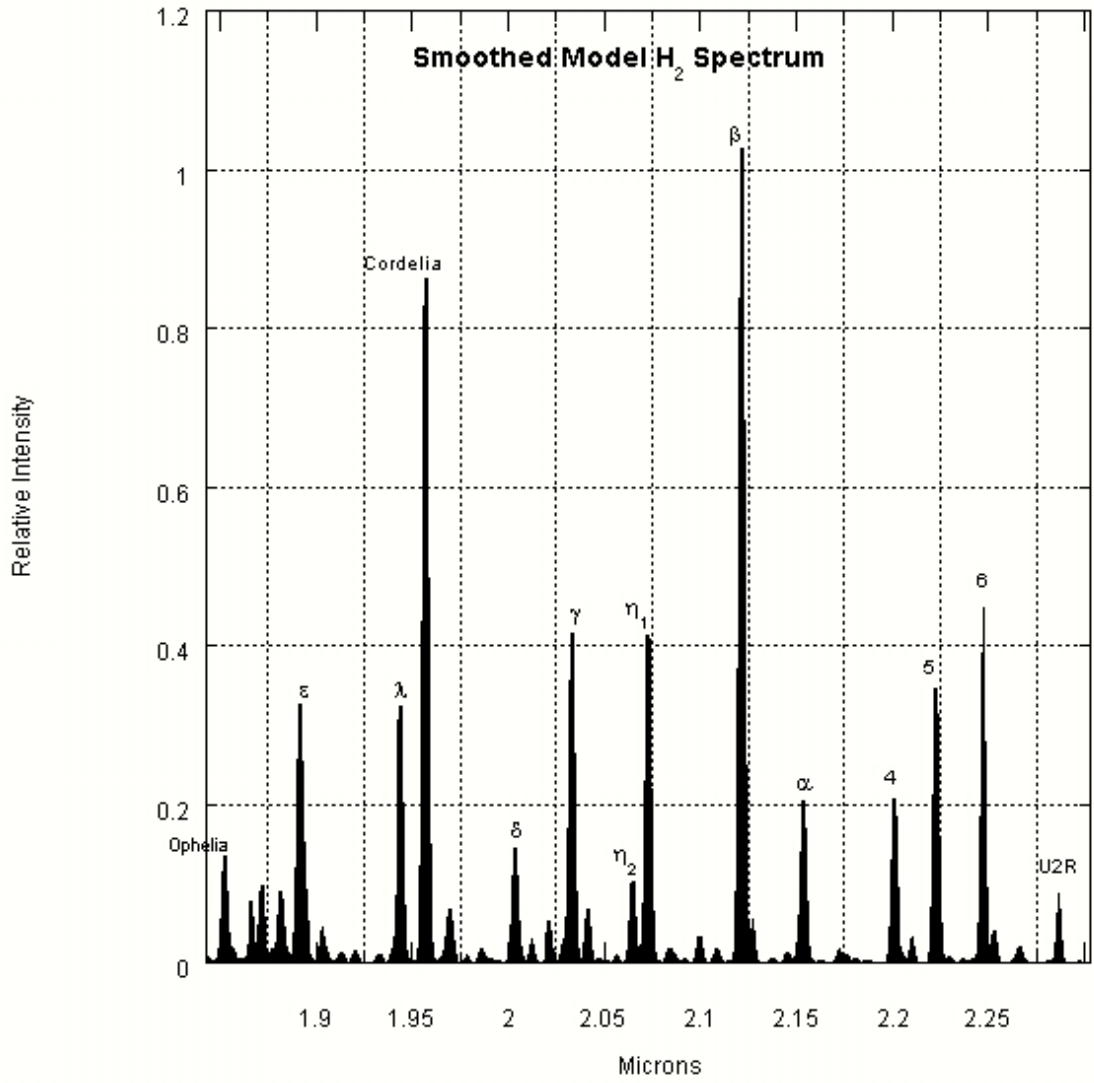

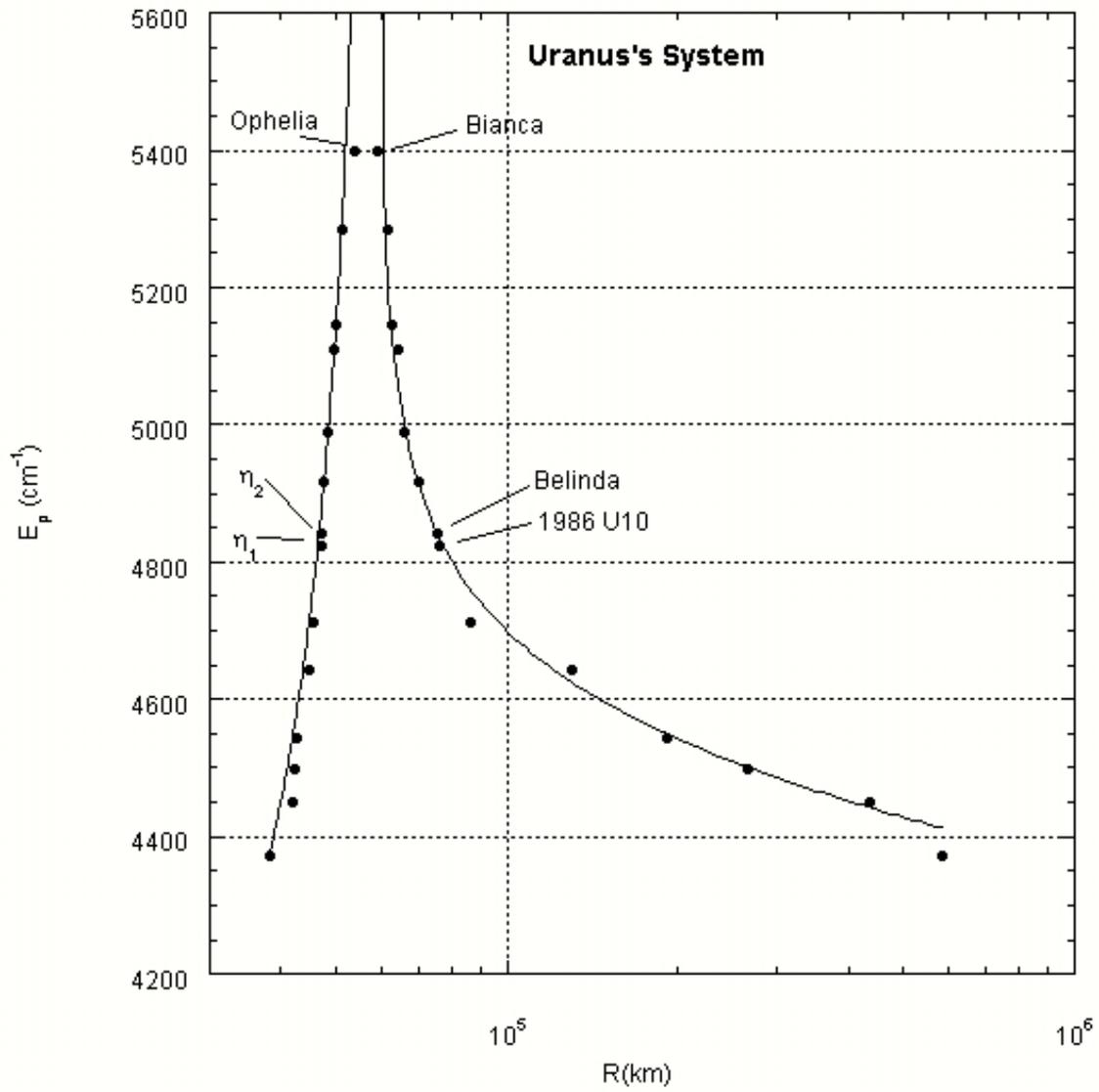

**Figure 5**

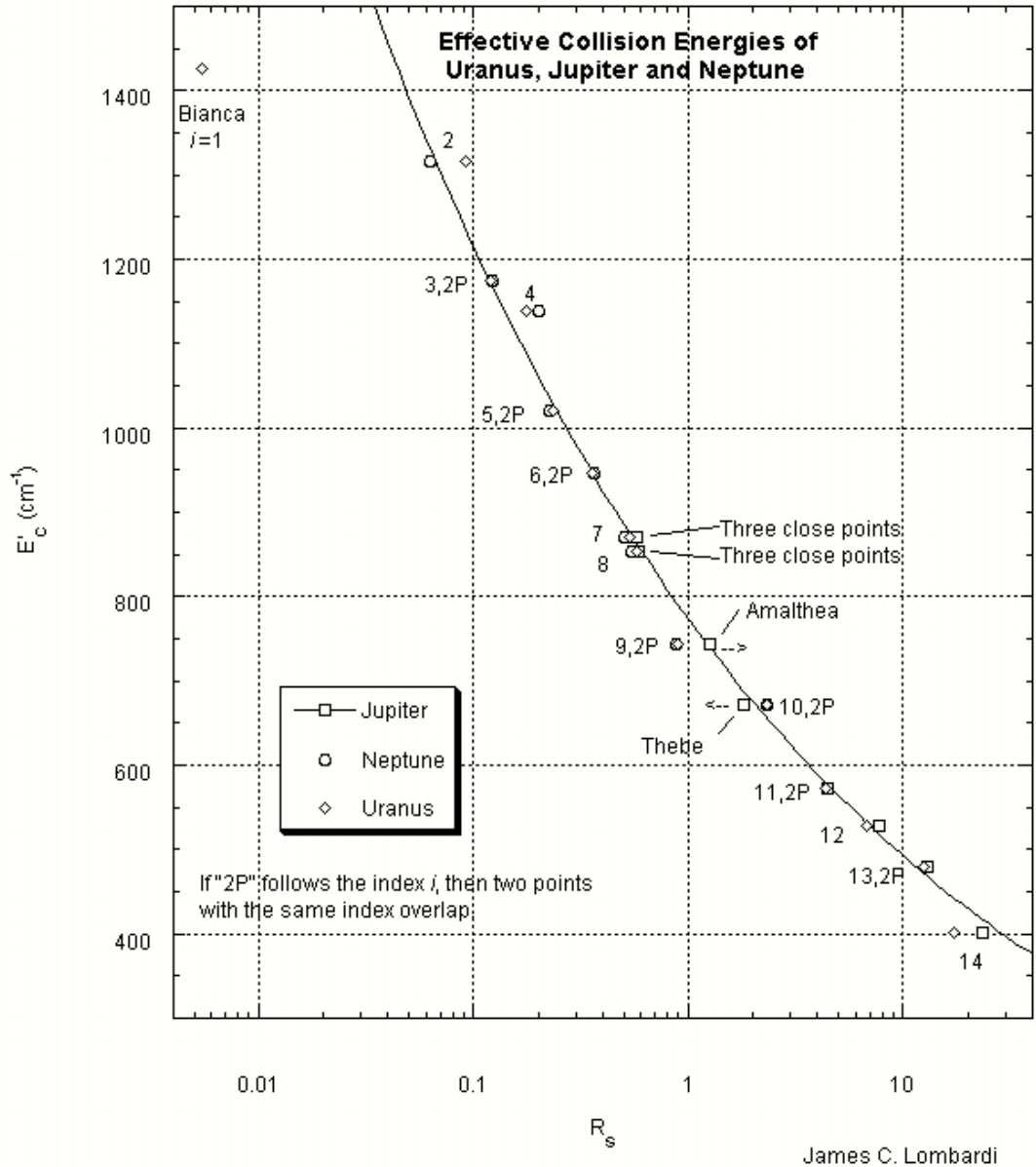

**Figure 6**

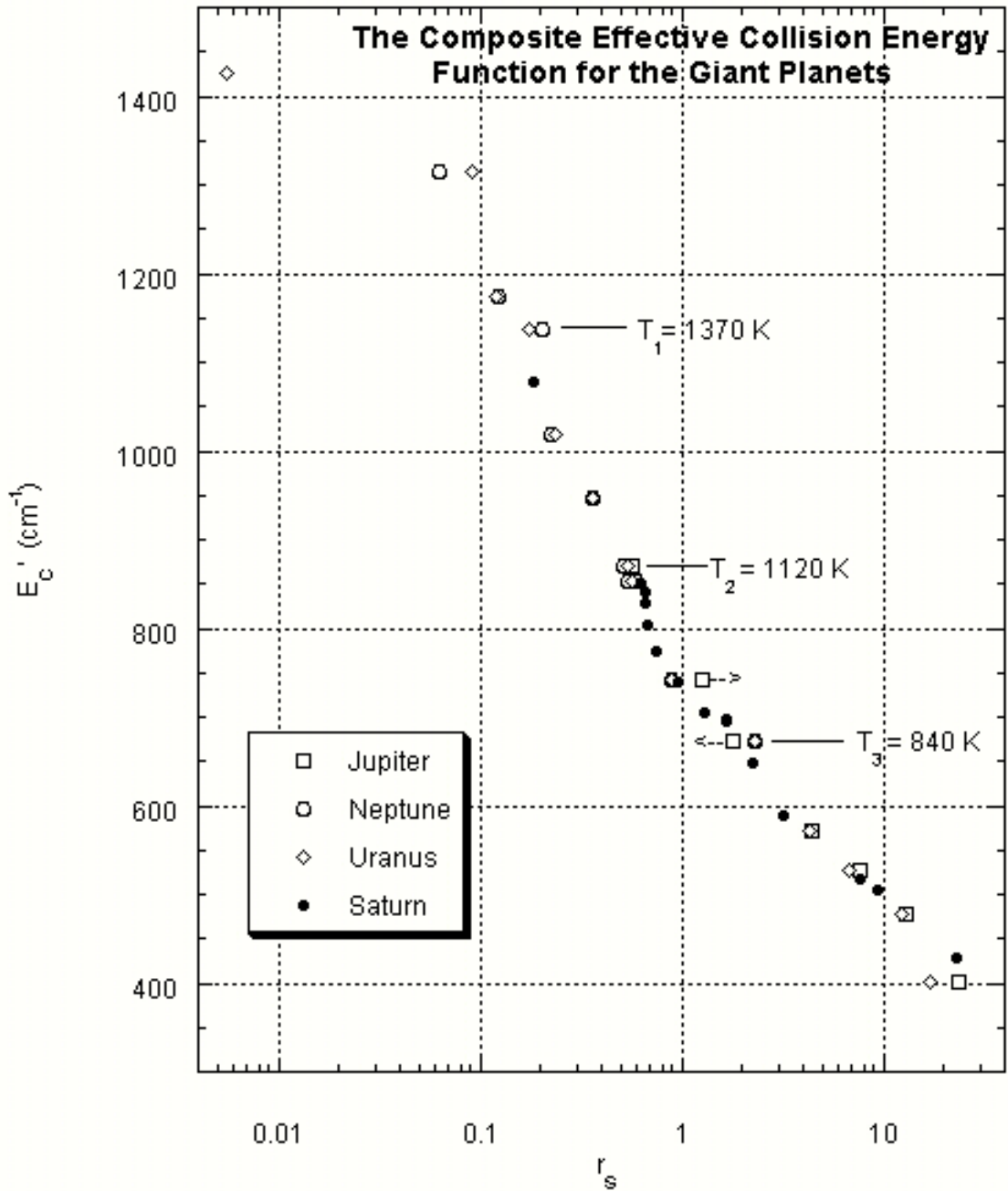

James C. Lombardi

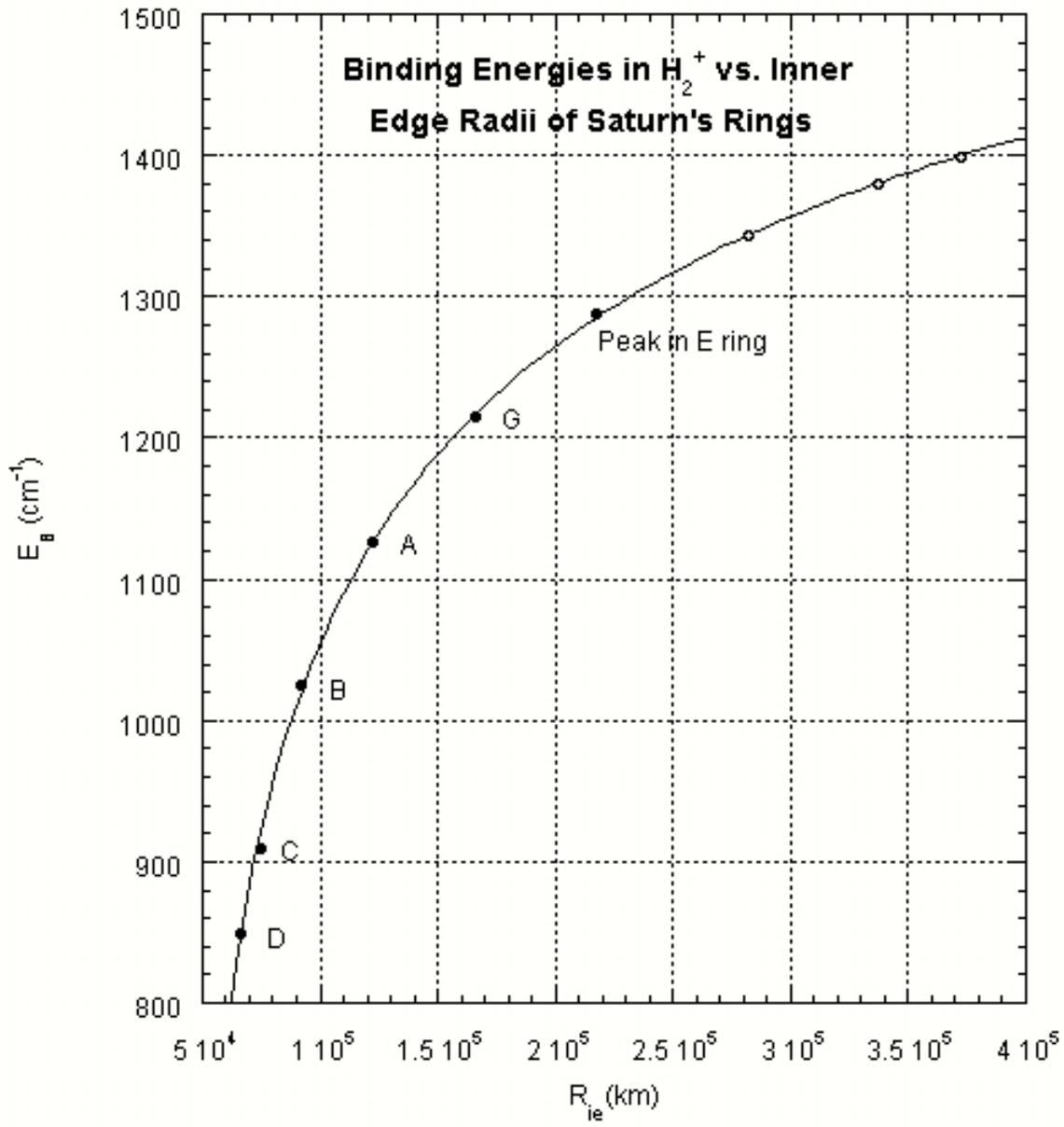

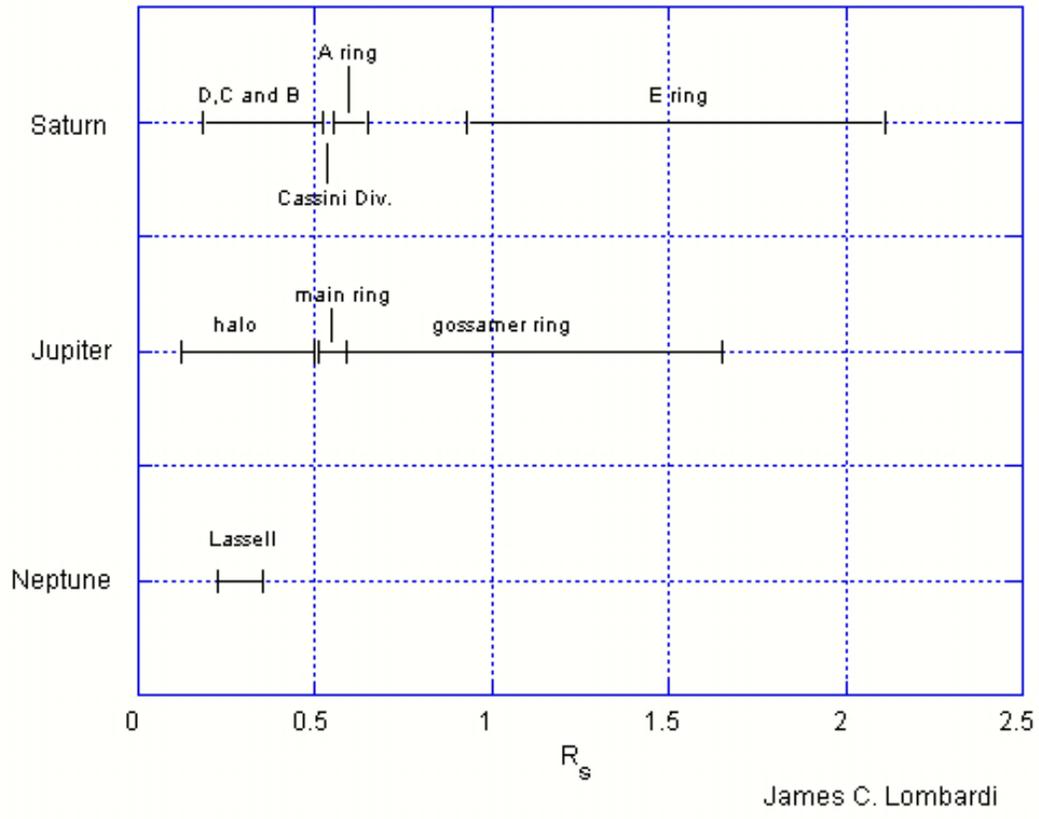

**Figure 8**